\documentclass[aps,prl,reprint,superscriptaddress]{revtex4-2}
\usepackage{amssymb}
\usepackage{amsmath, bm}
\usepackage{graphicx, color}
\usepackage[dvipsnames]{xcolor}
\usepackage{bbm}
\usepackage[bottom]{footmisc}
\usepackage{multirow}
\usepackage{soul}

\usepackage[normalem]{ulem}
\usepackage{xpatch}
\makeatletter
\xpatchcmd{\@ssect@ltx}{\@xsect}{\protected@edef\@currentlabelname{#8}\@xsect}{}{}
\xpatchcmd{\@sect@ltx}{\@xsect}{\protected@edef\@currentlabelname{#8}\@xsect}{}{}
\makeatother
\usepackage{hyperref}
\hypersetup{
    pdftoolbar=true,
    pdfmenubar=true,
    bookmarksopen=true,
    bookmarksnumbered=true,
    breaklinks=true,
    linktocpage,
    colorlinks=true,
    linkcolor={blue!95!black},
    citecolor={blue},
    urlcolor={blue!95!black},
}

\begin{document}
\title{Tunable shear thickening in active frictional non-Brownian suspensions}
\author{Bhanu Prasad Bhowmik}
\email{bhowmikbhanuprasad592@gmail.com}
\affiliation{Department of Physics, Indian Institute of Science Education and Research Pune, Pune-411008, India.}
\affiliation{School of Engineering, University of Edinburgh, Edinburgh EH9 3JL, United Kingdom}

\author{Christopher Ness}
\affiliation{School of Engineering, University of Edinburgh, Edinburgh EH9 3JL, United Kingdom}

\begin{abstract} 
 We study tunable shear thickening in active suspensions of non-Brownian, repulsive, frictional grains using particle-based simulation,
finding that activity augments the rheology beyond the friction-mediated shear thickening paradigm. Specifically, up to a crossover activity strength, increasing particle self-propulsion drives a viscosity-reducing `dethickening' of the system at large stresses, where the material would otherwise be in a thickened, highly viscous state. Above this crossover activity strength, the system enters a thickening regime.
Self-propulsion introduces additional isotropic dynamics to the particles,
which compete with the flow-driven formation of frictional contacts.
The degree of dethickening can thus be tuned by varying a suitably-defined dimensionless active stress that quantifies this competition.
Recognising the parallels between self-propulsion and other contemporary routes to dethickening, 
we demonstrate that our data, within the dethickening regime, are consistent with a recently proposed universal scaling framework, offering a route to tuning the rheology of dense suspensions. Moreover, using the obtained scaling relation, we infer how the shear-jamming and shear-thickening regimes can be tuned by choosing an appropriate self-propulsion strength, and partially validate these inferences with targeted simulations.
\end{abstract}

\maketitle

\paragraph{Introduction.}
Flows of dense suspensions and slurries are frequently observed in industry and natural processes~\cite{Guazzelli_Pouliquen_2018,NessReview}.
One widespread rheological phenomenon is shear thickening,
where the relative viscosity of the suspension $\eta_r$ increases with flow rate $\dot{\gamma}$ or applied stress $\sigma$~\cite{ShearThickening1958,shearThickening1989,
shearThickening1996, shearThickening2004, Brown2010}.
There is a well-established consensus that this behaviour originates in the stress-mediated onset of frictional contacts between particles~\cite{SetoMariPRL2013, MariJoR2014,RomainMariPNAS2015,MariSinghJoR2018}.
For a solid volume fraction $\phi$ sufficiently smaller than the jamming volume fraction $\phi_J$,
with increasing $\sigma$ or $\dot{\gamma}$ the system exhibits a smooth transition between a low-$\eta_r$ lubricated state,
where repulsive forces separate particles,
and a high-$\eta_r$ frictional state.
For $\phi$ close to $\phi_J$ the transition occurs abruptly with a discontinuous jump appearing in the constitutive curve $\eta_r(\dot{\gamma})$~\cite{DST2014WyartCates,MRamaswamyJoR2023, MRamaswamyPRL2025,barth2025universalscalingshearthickening}.  

Over the decades, shear thickening has drawn the attention of rheologists from both fundamental and applied domains.
In the former, recent studies have examined the distinction between isotropic jamming and the stress-induced shear-thickening transition that, at sufficiently high volume fraction, can culminate in shear jamming~\cite{SetoMariPRL2013, DST2014WyartCates}. Recent work~\cite{MorrisChakraborty2022, Sarika2022, MRamaswamyJoR2023, MRamaswamyPRL2025, barth2025universalscalingshearthickening} shows that both are associated with a critical transition but likely with different critical exponents.
From the applied point of view,
 large increases in viscosity due to thickening can pose challenges in the flow of the suspensions such as pipe blockages and extremely sluggish flow of materials in food processing industries.
Understanding and developing new ways to control, tune and exploit shear thickening behaviour is thus crucial and is beginning to receive much attention~\cite{Blanc_Lemaire_Peters_2014, PRR_TunableThick,HRZS_Isa_2018_PNAS,LNCSC_PNAS_2016,James2019, NMC_ScAdv_2018}.

The key goal when addressing tunable shear thickening is to implement interventions that manipulate and, ideally, remove, the frictional contacts present at high $\sigma$ or $\dot{\gamma}$ during flow.
Although in practice this is typically approached as a formulation challenge, \emph{i.e.} by adding surfactants or other molecules that control the physics of the particle-particle interactions~\cite{HRZS_Isa_2018_PNAS, FernandezPRL2013, PRR_TunableThick, James2019}, recently several methods have been proposed to achieve this \emph{on-the-fly}, that is, to impose a mechanical disturbance during flow that can affect the microstructure without changing the composition.
This has been demonstrated by both
orthogonal superposition (OSP) flow~\cite{LNCSC_PNAS_2016,NMC_ScAdv_2018}
and by targeted use of vibrations~\cite{hanotin2012vibration,garat2022using} or
acoustic perturbations (AP)~\cite{Mramaswamy_AP_PRL2019,SRO_Ness_CK_PRR_2024}.
In AP,
the suspension is subjected to acoustic noise, which effectively introduces repulsive interactions among the particles~\cite{acousticRepulsiveBrussPRE2012,acousticRepulsiveBrussPRE2013, acousticRepulsiveAiPRL2018} that prevent frictional contact formation.
In the case of OSP,
cyclic shear is applied orthogonal to the primary flow direction,
that is, a flow in direction $x$ and velocity gradient in $y$ would be supplemented by oscillatory shear applied
with flow along $z$ and velocity gradient in $y$.
Doing so,
with sufficiently small amplitude,
breaks the fragile load-bearing force network formed by simple shear alone,
again preventing frictional contacts from forming.
Interestingly, recent studies~\cite{bhowmik2026PRE, PeshkovEricClement2016} suggest that the inclusion of a small number of active particles into an otherwise conventional shear thickening dense suspension
can, dependent on the strength of activity,
produce a low viscosity state even at large $\dot{\gamma}$ through activity-induced contact suppression.
Analogous, perhaps, to the orthogonal cross shear mentioned above,
the self-propulsion of the active particles contributes additional dynamics to the non-Brownian system that apparently disrupt the formation of a force chain network.
By controlling the strength of activity, one can control the presence of frictional contacts and thus the viscosity.
This observation suggests self-propulsion as a suitable candidate to achieve tunable shear thickening.   

Here we use particle-based simulation,
namely a variant of the discrete element method,
to study tunable shear thickening of active dense suspensions of non-Brownian, frictional particles in the presence of repulsive interactions.
The active particles are modelled as non-Brownian run-and-tumble particles~\cite{clementPRL2015,Clement2016,RituparnoSoftMatter},
characterised by an active driving force $f^a$ and a persistence time $\tau_p$.
We find that shear thickening can be tuned by varying the strength of the active forcing
relative to the repulsive force acting between particles.
The isotropic self-propelling motion of the active particles effectively prevents the formation of shear-induced force-chain networks and reduces the number of frictional contacts in agreement with Ref.~\cite{bhowmik2026PRE}.
We then compare this mechanism with the tunable shear thickening achieved by applying cyclic shear orthogonal to the primary flow direction, where force chains break due to the additional particle motion induced by the cyclic forcing.
We show that tunable shear thickening with activity can be described using the same universal scaling framework proposed for cyclic perturbations, provided all parameters are mapped appropriately.
Finally, we discuss how the discontinuous shear thickening (DST) and shear jamming (SJ) region in the $\sigma^*$–$\phi$ phase diagram decreases in size with increasing activity strength.

\paragraph{Simulation details.}
 We simulate $N = 10^3$ non-Brownian
particles with Coulombic friction coefficient $\mu_p = 1$,
suspended in a density $\rho$ matched fluid of viscosity $\eta_f$.
The particles have two different radii, $a$ and $1.4a$, mixed in equal numbers and placed in a periodic, cubic box of length $L$.
The rheological response is obtained by imposing a velocity profile $\bm{U}^\infty(y) = \dot{\gamma} y \hat{\bm x}$ to the suspending fluid phase,
in the presence of Lees-Edwards boundary conditions across $y$.
Here $y$ is the distance along the direction of the velocity gradient and $\hat{\bm x}$ is the unit vector along the flow direction.

A particle with radius $a_i$ located at position ${\bm r}$ experiences a viscous drag force
$\bm{f}_i^d = 6\pi\eta_f a_i\left(\bm U^{\infty}(r_y) - \bm u_i \right)$
and a torque
$\bm \tau _i^d = 8\pi\eta_f a_i^3\left(\bm{\Omega}^{\infty} - \bm{\omega}_i \right)$.
Here $\bm{u}_i$ and $\bm{\omega}_i$ are the linear and angular velocities of the particle,
and
$\bm{\Omega}^{\infty} = \frac{1}{2} \left( \nabla \times \bm{U}^{\infty} \right)$.
Additionally,
the relative velocity of a pair of particles $\bm{u}_{ij} = \bm{u}_j - \bm{u}_i$ is resisted by the viscous fluid,
modelled here as short range hydrodynamic interactions~\cite{Jeffrey_Onishi_1984, KimAndKarila, ChealAndNess}.
The leading order term of this lubrication force is given by
$\bm f_{ij}^{h}  \sim \frac{\eta_f}{\delta_{ij}}\left(\bm{u}_{ij} \cdot \bm{n}_{ij}\right)\bm{n}_{ij}$,
while that of the torque on the $i^{th}$ particle due to the interaction with $j^{th}$ particle is given by
$\bm {\tau}_{ij}^{h}   \sim \ln \left( \frac{a_i\eta_f}{\delta_{ij}} \right) \left(\bm{u}_{ij} \times \bm{n}_{ij} \right)$.
Here $\delta_{ij} = |\bm{r}_j - \bm{r}_i|-(a_i + a_j)$ is the separation between the particle surfaces,
and $\bm{n}_{ij} = \left(\bm{r}_j - \bm{r}_i\right)/|\bm{r}_j - \bm{r}_i|$ is the center-to-center unit vector.
We use $\delta_{ij} = 0.001a'$ in the force calculation whenever the measured $\delta_{ij} \leq 0.001a'$,
allowing particles to come into direct contact.
Here, $a'$ is the radius of the smaller particle.
Additionally, we consider lubrication interactions only for pairs that have $\delta_{ij} < 0.5a'$, to reduce computational expense.

When $\delta_{ij} < 0$ the particles experience frictional contact forces, modelled as history dependent Hookean interactions.
The normal component of the contact force is given by
$\bm f_{ij}^{c,n}  = - k_n \delta_{ij}\bm{n}_{ij}$
and the tangential component is given by
$\bm f_{ij}^{c,t}  =  - k_t \bm{\xi}_{ij}$.
Meanwhile the torque is given by $\bm {\tau}_{ij}^{c}  = a_i\left(\bm{n}_{ij} \times \bm f_{ij}^{c,t}\right)$.
Here
$k_n$
and
$k_t$
are the normal and tangential stiffnesses,
and $\bm{\xi}_{ij}$ denotes the accumulated tangential displacement between particles, measured from the moment they first come into contact until the contact breaks.
This displacement accounts for the history dependence of the frictional force~\cite{CundallAndStrackDEM}.
Friction introduces an additional constraint into the system, preventing sliding between two particles when the ratio of their tangential to normal interaction force is less than the friction coefficient.
Additionally, the particles repel each other via a screened electrostatic repulsion given by
$\bm f_{ij}^r  = - f^r \text{e}^{-r_{ij}/r^*} \bm{n}_{ij}$
when
$\delta_{ij} > 0$
and
$-f^r \text{e}^{-(a_i + a_j)/r^*}\bm{n}_{ij}$
for
$\delta_{ij} < 0$.

Finally,
in addition to these interactions we introduce a randomly oriented force
$\bm{f}_i^a = f^a \bm{e}_i$,
applied independently on every particle to produce the self-propulsion,
enforcing the constraint that $\sum_{i=1}^{N}\bm{f}_i^a = 0$ so that the center of mass does not drift.
The orientation vector is constructed from discrete Cartesian components as $\bm{e}_i = \frac{1}{\sqrt{3}} \left(e_x\,\hat{e}_x + e_y\,\hat{e}_y + e_z\,\hat{e}_z \right)$. Here, $\hat{e}_x$, $\hat{e}_y$, and $\hat{e}_z$ denote the unit vectors along the $x$-, $y$-, and $z$-axes, respectively, of the laboratory Cartesian coordinate system.
Each component $e_x$, $e_y$, and $e_z$ independently takes values from the discrete set $\{-1, 1\}$. 
After a persistence time interval $\tau_p$, a new orientation is assigned by independently drawing the Cartesian components again from the same discrete distribution~\citep{RituparnoSoftMatter, RishabSmarajitNatPhy2025}.
Motivated by OSP results in which the dethickening effect is reported to saturate at large frequencies,
we focus our exploration of the parameter space by setting $\tau_p\dot{\gamma}\sim 10^{-3}$ throughout the following.
We then define two dimensionless control parameters $\sigma^*= \sigma_{xy}a^2/f^r$
and
$\sigma^*_a = f^a/6\pi\eta_f\dot{\gamma}a^2$
that compare,
respectively,
the hydrodynamic stress to the repulsive stress and the active stress to the hydrodynamic stress.
Here $\sigma_{xy}$ is the shear stress measured in our simulations.
Notably,
we choose to construct these so that
increasing $\sigma^*$ leads to shear thickening
whereas
increasing $\sigma^*_a$ leads to dethickening.
(In practice this is achieved by varying $f^r$ and $f^a$, having verified that our results are invariant under changes in $\dot{\gamma}$.) We also simulate our system with $N = 10^4$ and a continuous distribution of active forces to show that no significant finite size effect is observed in the presented data and that, in the limit of the chosen persistence time, the choice of a discrete active force results in a similar rheological response to a continuous distribution.
\begin{figure}
\includegraphics[width=0.48\textwidth]{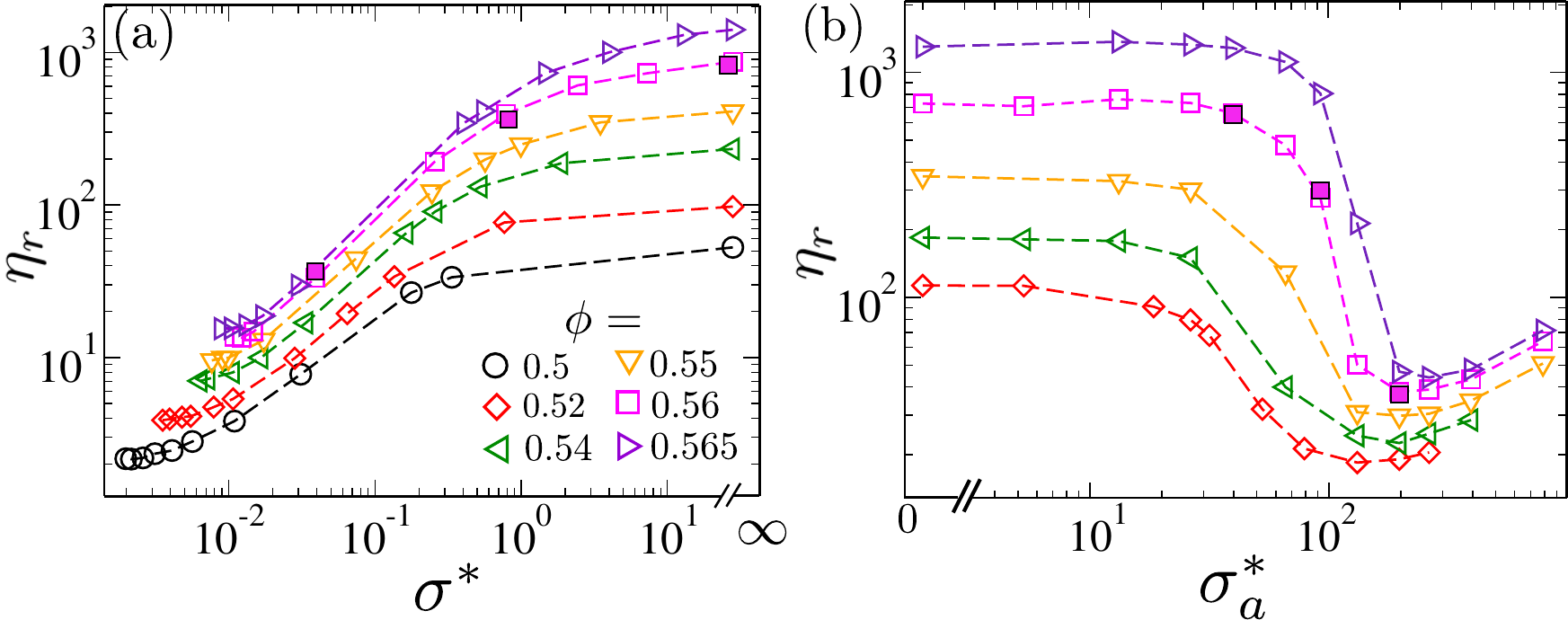}
\caption{Relative viscosity of an active shear thickening suspension.
Shown are the variations of relative viscosity $\eta_r$ as functions of
(a) the dimensionless applied stress $\sigma^{*} = \sigma_{xy}a^2/f^r$ for fixed active stress $\sigma^*_a = 0$
and
(b) the dimensionless active stress $\sigma_a^{*} = f^a/6\pi\eta_f\dot{\gamma}a^2$
 for a range of volume fractions $\phi$ and for fixed applied stress $\sigma^*=7.29$.
 The legend in (a) applies also to (b). The solid symbols in (a) show data for $N = 10^4$, and the solid symbols in (b) show data for a continuous distribution of active forces.
}
\label{fig1}
\end{figure}
\begin{figure*}
\includegraphics[width=0.98\textwidth]{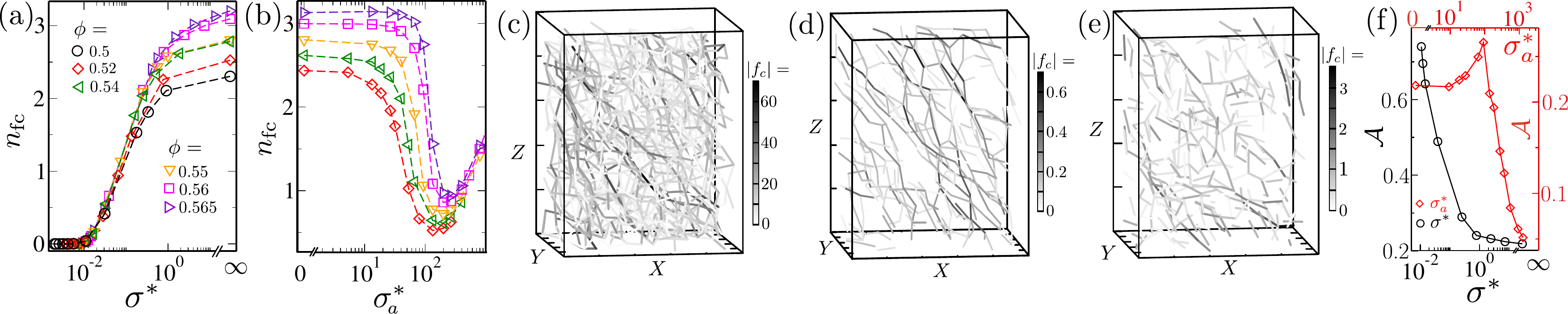}
\caption{Microstructure of an active shear thickening suspension.
Shown are the variations of the mean number of frictional contacts $n_\mathrm{fc}$ as a function of (a) the dimensionless stress $\sigma^*$ and (b) the dimensionless active stress $\sigma_a^*$.
(c) Shows an example force chain network taken from a typical configuration with $\phi = 0.56$ in the steady state with $\sigma^* = \infty$ and $\sigma_a^* = 0$.
(d) and (e) show the same as (c), but now with $\sigma^* = 0.065$, $\sigma_a^* = 0$, and  $\sigma^* = 7.29$, $\sigma_a^* = 198.95$ respectively, which have same $n_\mathrm{fc} \sim 1$. The colour bar stands for the magnitude of the contact force.
(f) shows the variation of the anisotropy factor $\mathcal{A}$ with both $\sigma^*$ (black circles) and $\sigma_a^*$ (red diamonds).}
\label{fig2}
\end{figure*}
\paragraph{Results.}
First we study systematically how the dimensionless stresses $\sigma^*$ and $\sigma_a^*$ control the viscosity.
Fig.~\ref{fig1}(a) shows the dependence of the relative viscosity $\eta_r = \sigma_{xy}/\dot{\gamma}\eta_f$
 on $\sigma^*$ for a range of solid volume fraction $\phi$ and a fixed value of $\sigma^*_a=0$.
The result reflects the standard behaviour of a shear thickening non-Brownian suspension of passive particles.
Specifically,
at small $\sigma^*$ the particle interactions are lubrication dominated due to the repulsive force.
As $\sigma^*$ increases so does the number of direct mechanical contacts,
resulting in a smooth increase in viscosity.
When $\sigma^*$ exceeds a $\phi$-dependent critical value,
the applied stress overcomes the pairwise repulsive barrier set by $f^r$, leading to a rapid increase in $\eta_r$.
The slope of the $\eta_r-\sigma^*$ curve increases with $\phi$ and approaches unity for the largest studied $\phi$ which indicates discontinuous shear thickening.

In Fig.~\ref{fig1}(b) the equivalent dependence is shown for $\sigma_a^*$, measured at a fixed $\sigma^* = 7.29$.
Here $\sigma^*$ is large enough that the rheology is dominated by the frictional contacts for the passive system,
thus producing a large,
shear thickened,
viscosity at $\sigma^*_a = 0$.

Now,
as $\sigma_a^*$ increases, the viscosity varies systematically.
First, at small $\sigma_a^*$, the viscosity remains unchanged indicating that the strength of the self-propulsion stress is too low to affect the overall dynamics of the system.
As $\sigma_a^*$ increases, the viscosity decreases abruptly,
primarily due to prevention of shear induced force chain formation as we will see in the next section.
The viscosity then reaches a minimal value at a $\phi$-dependent value of $\sigma^*_a$,
beyond which there is then growth of $\eta_r$ with $\sigma_a^*$. The reason behind such an increase in viscosity at large $\sigma_a^*$ is the increase in isotropic frictional contacts due to the activity-induced diffusion of particles, as explained in Ref.~\cite{bhowmik2026PRE}.
This trend remains for all the studied $\phi$,
with both the absolute and relative viscosity drops increasing as $\phi$ approaches $\phi_J$.
Overall the simulation data clearly show the opportunity for tunable shear thickening by varying $\sigma_a^*$,
and raise the fundamental question of whether there is a microscopic connection between dethickening achieved using self-propulsion as done here, and by mechanical perturbation along the direction orthogonal to shear flow~\cite{NMC_ScAdv_2018}.

We next seek to understand the microstructural changes that take place in the presence of self-propulsion,
and how these lead to tunable shear thickening.
To do so, we focus on the quantity $n_\mathrm{fc}$, enumerating the number of frictional contacts per particle averaged over the system
and representing the primary microscopic origin of the high viscosity state and jamming.
The dependence of $n_\mathrm{fc}$ on $\sigma^*$ and $\sigma_a^*$ is shown in Fig.~\ref{fig2}.
As $\sigma^*$ increases at $\sigma^*_a = 0$,
$n_\mathrm{fc}$ increases smoothly from close to zero, its slope increasing modestly with increasing $\phi$, Fig.~\ref{fig2}(a).
Again, this is consistent with the prevailing picture of shear thickening, in which frictional contacts become prevalent as a repulsive force is overcome at large stress.
Meanwhile the dependence on $\sigma_a^*$ follows the opposite trend, Fig.~\ref{fig2}(b),
suggesting a competition between the applied and active stresses.
While the external driving leads to the proliferation of contacts by overcoming the repulsive barrier $f^r$,
the self-propulsion reduces the number of contacts due to its additional isotropic motion,
thus resulting in a low viscosity state at a $\phi$ which would otherwise be highly viscous. Such variations in the fraction of contacts were studied in Ref.~\cite{bhowmik2026PRE}. Our findings, with the addition of a repulsive potential, are consistent with the previously reported results.
The evidence of such an effect is clear in the force chain network,
shown in Figs.~\ref{fig2}(c)--(e) for three different conditions,
namely large (c) and small (d) $\sigma^*$ in the absence of activity ($\sigma^*_a = 0$),
and (e) with $\sigma^*$ and $\sigma^*_a$ both large. (d) and (e) have the same number of frictional contacts with $n_\mathrm{fc} \sim 1$.
When the applied stress is strong and active stress is weak or absent,
the particles come into mechanical contact and form anisotropic force chains that span the system as shown in Fig.~\ref{fig2}(c).
At small applied stress, the contacts are more lubrication dominated with a smaller force scale, as shown by the colour map, however, they still retain the highly anisotropic structure (Fig.~\ref{fig2}(d)).
Similarly, when the high active stress dominates, force chain network formation is prevented, leading to more isolated, pairwise contacts and consequently a lower viscosity (Fig.~\ref{fig2}(e)). 
To quantify the nature of the force-chain network, we calculate the anisotropy factor from the force-weighted fabric tensor, $Q_{\alpha\beta}^f =
\frac{\sum_{i=1}^{N_c} f_n^i n_\alpha^i n_\beta^i}
{\sum_{i=1}^{N_c} f_n^i}$, following Refs.~\cite{GUOFabricTensor, ShaebaniFabricTensor, PandareFabricTensor}. Here, $N_c$ is the total number of contacts, $f_n^i$ is the magnitude of the normal force between the two particles forming the $i$th contact, and $\bm{n}^i$ is the unit vector joining their centres. The anisotropy factor is defined as $\mathcal{A} = \sqrt{\frac{3}{2} \left(Q_{\alpha\beta}^f-\frac{1}{3}\delta_{\alpha\beta}\right)
\left(Q_{\alpha\beta}^f-\frac{1}{3}\delta_{\alpha\beta}\right)}$, where summation over repeated Cartesian indices is implied. With this definition, $\mathcal{A}=0$ for an isotropic force-chain network. The variation of $\mathcal{A}$ with $\sigma^*$ and $\sigma_a^*$ is shown in Fig.~\ref{fig2}(f). As $\sigma^*$ increases, the number of branches decreases, leaving only a few branches that bear most of the load. This makes the force-chain network more anisotropic, as also supported by Fig.~\ref{fig2}(c). However, $\mathcal{A}$ exhibits a non-monotonic dependence on the active stress. At small $\sigma_a^*$, activity breaks frictional contacts, thereby reducing the number of branches and increasing the anisotropy factor, similar to the effect of increasing the applied stress. At large $\sigma_a^*$, activity dominates over the anisotropic driving and induces activity-driven diffusion, resulting in a more isotropic frictional contact network, as indicated by the sharp decrease in $\mathcal{A}$.
\begin{figure*}
\includegraphics[width=0.98\textwidth]{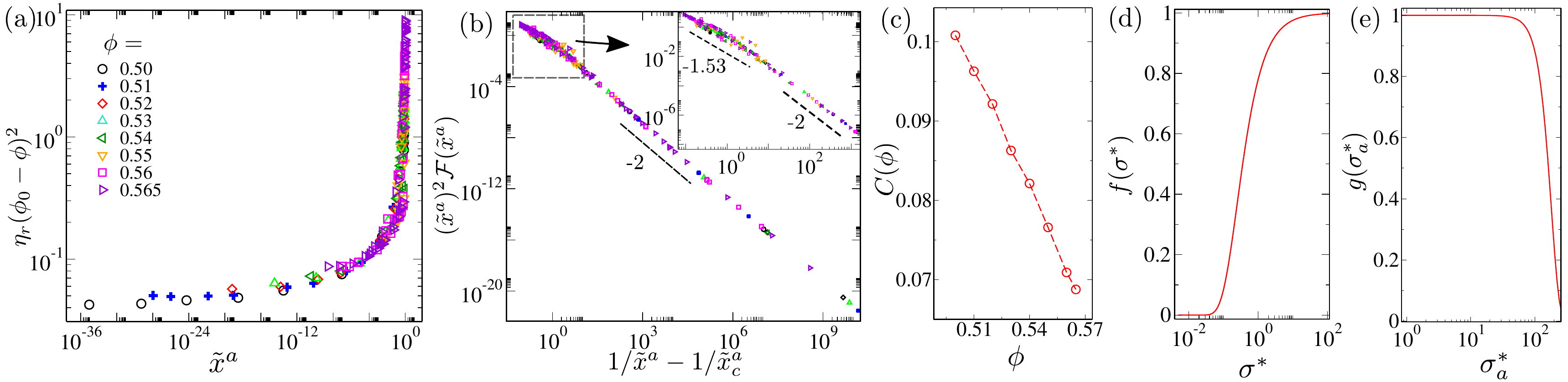}
\caption{Scaling of the rheology data for active shear thickening suspensions. Shown in (a) is a collapse of the data from Figs.~\ref{fig1}(a) and (b) for a range of $\phi$, $\sigma^*$ and $\sigma^*_a$, onto a single master curve with the scaling variable for our active system being $\tilde{x}^a = f(\sigma^*)C(\phi)g({\sigma_a}^*)/(\phi_0 - \phi)$.
(b) Two power law regimes, with exponent $-2$ associated with isotropic jamming and exponent $-1.53 \pm 0.03$ (inset) associated with friction-dominated jamming.
(c) Variation with volume fraction $\phi$ of the scaled fabric anisotropy factor $C(\phi) = 0.33 \mathcal{A}(\phi)$;
sketches of the functional forms (d) $f(\sigma^*)$ and (e) $g(\sigma_a^*)$ used in the scaling variable for the data collapse. 
}
\label{fig3}
\end{figure*}

Tunable shear thickening has been achieved previously by various methods.
Two standard ones are the use of acoustic perturbation to the flowing samples,
and applying oscillatory mechanical perturbation orthogonal to the primary flow direction.
The acoustic perturbation can be understood by considering generation of effective repulsive potentials.
In the case of orthogonal perturbation,
the motion of the particles out of the shear plane reduces the number of frictional contacts,
an effect that grows with the amplitude of the perturbation and oscillation frequency.
Here the mean number of frictional contacts first decreases until a minimum value is achieved,
before then increasing due to inertial motion~\cite{LNCSC_PNAS_2016,NMC_ScAdv_2018}.
The additional dynamics of the particles due to cross shear is analogous to what we observe in our active system,
the difference being that in one case the additional particle dynamics are due to external driving,
whereas in the other they are due to self-propulsion.
Direct comparison suggests that the amplitude of the active force $f^a$ plays a similar role as the amplitude of the oscillation $\gamma_\mathrm{OSP}$,
while the persistence time $\tau_p$ has the same role as the inverse oscillation frequency $\omega_{\mathrm{OSP}}$.
Recent studies in active amorphous solids also suggest that cyclic shear and the presence of self-propelled particles with finite persistence time result in similar rheological responses~\cite{RishabSmarajitNatPhy2025, Goswami2025}.

We further explore this analogy by testing the validity of a proposed scaling framework
that seeks to universally describe the tunable shear thickening obtained under mechanical or acoustic forcing
by collapsing the vast rheology data onto a single master curve.
The scaling function in this data collapse shows a crossover between two critical points;
a lubrication dominated divergence of viscosity at random close packing $\phi_0 \sim \phi_{\text{RCP}}$ to a friction dominated viscosity divergence at a lower volume fraction $\phi_J$,
as a function of a scaling variable defined as $\tilde{x} = f(\sigma)C(\phi)g(\dot{\Gamma})/(\phi_0 - \phi)$.
Here, $C(\phi)$ is a so-called anisotropy factor first introduced in Ref.~\cite{MRamaswamyJoR2023}, which exhibits a nonmonotonic dependence on $\phi$, with a maximum at an intermediate value.
$\dot{\Gamma}$ is a dimensionless quantity defined as $\gamma_{\mathrm{OSP}}\omega_{\mathrm{OSP}}/\dot{\gamma}$.
The functions $f(\sigma)$ and $g(\dot{\Gamma})$ take the form $f(\sigma) = \mathrm{e}^{-\left(\sigma_0/\sigma\right)^{0.75}}$ and $g(\dot{\Gamma}) = \mathrm{e}^{-\left(\dot{\Gamma}/\dot{\Gamma}_0\right)}$.
The scaling relation is given by $\eta(\phi_0 - \phi)^2 = \mathcal{F} (\tilde{x})$.
The scaling function $\mathcal{F}(\tilde{x})$ becomes a constant at small $\tilde{x}$,
corresponding to the divergence of the viscosity at $\phi_0$ with critical exponent $-2$ when the stress is low and interactions are lubrication dominated.
$\mathcal{F}(\tilde{x})$ diverges at a critical value $\tilde{x}_c$ corresponding to the friction dominated region occurring at high $\sigma$.


To implement this framework in our simulation setup,
we first replace the variables $\sigma$ by applied stress $\sigma^*= \sigma_{xy}a^2/f^r$ and $\dot{\Gamma}$ by active stress $\sigma^*_a = f^a/6\pi\eta_f\dot{\gamma}a^2$,
then define the scaling variable for the active system $\tilde{x}^a = f(\sigma^*)C(\phi)g(\sigma^*_a)/(\phi_0 - \phi)$.
We then aim to collapse all our data for a large set of $\phi$, $\sigma^*$ and $\sigma_a^*$, as shown in Fig.~\ref{fig3}(a), for a fixed persistence time $\tau_p$.
Doing so results in a reasonably good collapse of all the data for a large set of values of $\sigma^*$ and $\sigma_a^*$,
each for $\phi$ ranging from $0.5$ to $0.565$.
We note that in the case of active systems we can have the same value of $\eta_r$ for two different $\sigma_a^*$ due to the nonmonotonicity reported in Fig.~\ref{fig1}(b).
In these cases we considered data only for those values of $\sigma_a^*$ where the viscosity either remains constant or decreases with $\sigma^*_a$.
Our scaling function looks the same as that found in Ref.~\cite{MRamaswamyPRL2025},
asymptotically approaching a constant value at small $\tilde{x}^a$ and diverging at $\tilde{x}^a_c$.
Fig.~\ref{fig3}(b) shows the Cardy scaling~\cite{Cardy1996},
which manifests two different power law regimes.
Here the exponent $-2$ corresponds to jamming at the isotropic jamming point $\phi_0$,
while the exponent $-1.53 \pm 0.03$ corresponds to jamming due to friction dominated contacts at $\phi_J$. The data collapse involves three functions $C(\phi)$, $f(\sigma^*)$ and $g(\sigma_a^*)$, which are taken as fitting parameters in the previous studies~\cite{MRamaswamyJoR2023, MRamaswamyPRL2025, barth2025universalscalingshearthickening}. Here, we calculate the force-weighted anisotropy factor $\mathcal{A}$ for different values of $\phi$ at $\sigma^* = \infty$ and $\sigma_a^* = 0$, and use it to determine $C(\phi)$ through the relation $C(\phi) = \alpha \mathcal{A}(\phi)$. Here, $\alpha$ is a multiplicative constant that we tune to vary the range of $\tilde{x}^a$. We choose $\tilde{x}_c^a = 1$ and then tune $\alpha$ to obtain the power-law regime associated with the friction-dominated divergence of viscosity. We find $\alpha = 0.33$, and the resulting $C(\phi)$ decreases monotonically with $\phi$ within the studied range, as shown in Fig.~\ref{fig3}(c). We choose $f(\sigma^*) = e^{- \bar{\sigma}^*/\sigma^*}$ and $g(\sigma_a^*) = e^{-( \sigma_a^*/\bar{\sigma}_a^*)^{\nu}}$  as shown in Figs.~\ref{fig3}(d)--(e). The form of $f(\sigma^*)$ is motivated from the study in Ref.~\cite{MariSetoSoftMatter2019} where variation of frictional contacts due to applied stress is measured directly using a critical load model which shares rheological similarities with the model studied here. The compressed exponential form of $g(\sigma_a^*)$ is motivated by the observed decrease in the frictional contact number, $n_{\mathrm{fc}}$, with increasing $\sigma_a^*$ shown in Fig.~\ref{fig2}(b). Physically, it signifies the abrupt and cooperative collapse of frictional force chains when the active stress becomes comparable to the characteristic stress scale associated with the stability of the shear-induced contact network. In order to get the best data collapse we vary $\bar{\sigma}^*$, $\bar{\sigma}_a^*$ and $\nu$ and get $\bar{\sigma}^* = 0.25$, $\bar{\sigma}_a^* = 181.8$ and $\nu = 3.5$.
We note that the value of the obtained exponents is extremely sensitive to the value of $\tilde{x}^a_c$,
likely true also of the values reported from experiments. 
Notably, the exponent associated with the friction-dominated viscosity divergence is close to that reported in Ref.~\cite{MRamaswamyPRL2025}. The scaling analysis presented here involves several fitted parameters and should therefore be regarded as an empirically calibrated description consistent with the previously proposed framework, rather than a parameter-free demonstration of universality. The error associated with the exponent is solely related to the power law fit and does not represent the sensitivity associated with the choice of $\tilde{x}^{a}_c$ and other fitting parameters. Importantly, our system differs from the experimental system not least in that the frictional particles simulated here are modelled in a rather simplistic manner with several factors such as rolling friction and the hard-sphere limit not being taken into account.

\begin{figure}
\includegraphics[width=0.4\textwidth]{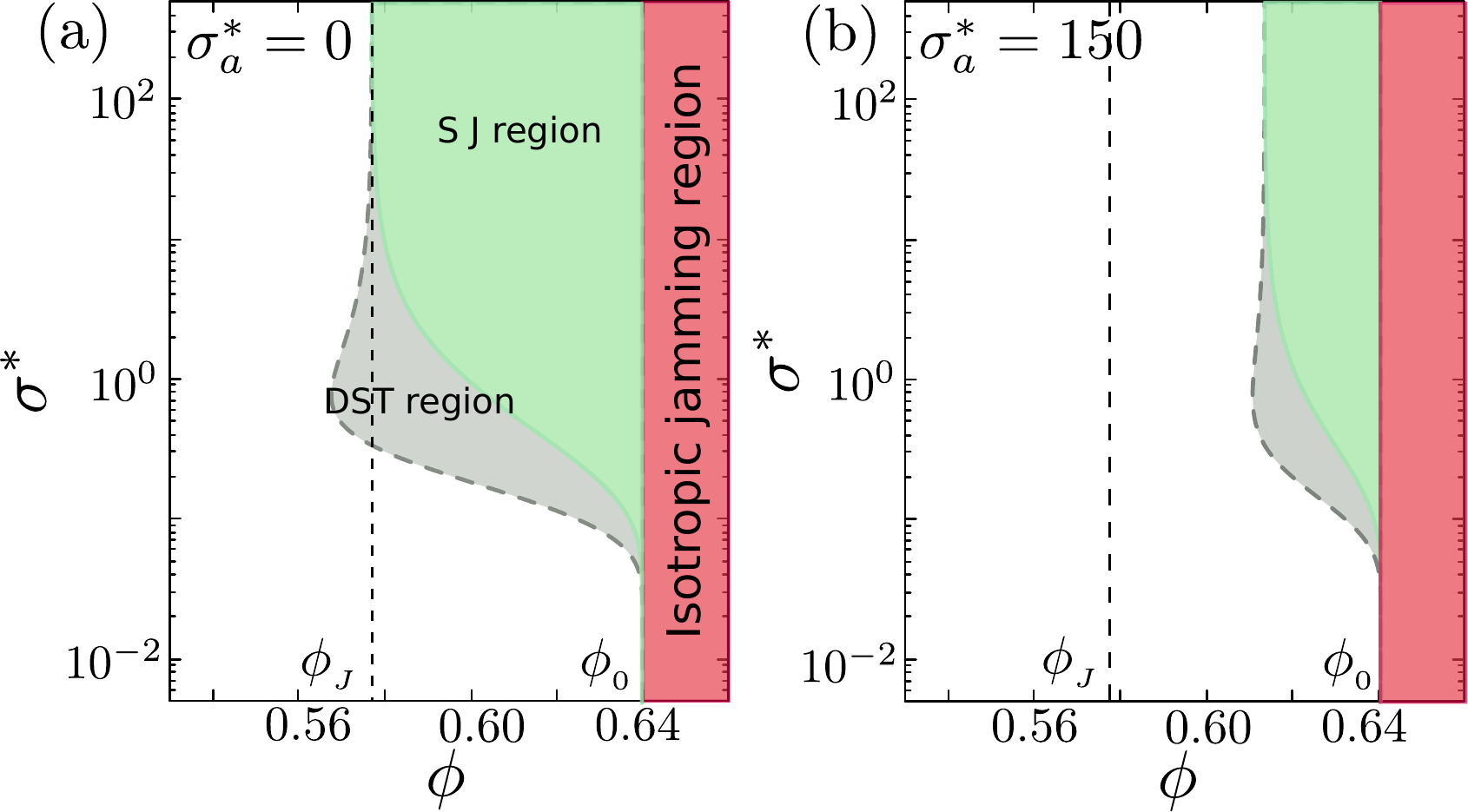}
\caption{Tuning of the shear jamming (SJ) and discontinuous shear thickening (DST) regions due to active stress.
(a) Isotropic jamming (red), shear jamming (green), and DST (gray) regions in the $\sigma^* - \phi$ phase diagram in the absence of activity, with $\sigma_a^* = 0$.
(b) Altered shear jamming and DST regions under a large active stress, $\sigma_a^* = 150$.}
\label{fig4}
\end{figure}

Finally, we discuss how the presence of activity alters the size of the shear thickening and shear jamming regions in the $\sigma^* - \phi$ phase diagram,
and thus offers a route to tuning the rheology.
In a typical shear thickening system of frictional granular particles,
 when $\phi > \phi_0$ the system exhibits isotropic jamming for any finite value of $\sigma^*$ as shown in red colour in Fig.~\ref{fig4}. Below $\phi_0$, there is a region (shown in green in Fig.~\ref{fig4}) where jamming occurs due to external shear when $\sigma^*$ is above a certain $\phi (> \phi_J)$ dependent value.
 There also exists a region at smaller applied stress (gray in Fig.~\ref{fig4}) where discontinuous shear thickening occurs.
 Following Ref.~\cite{MRamaswamyPRL2025}, we argue here that the boundary of the shear jamming region can be pushed towards the isotropic jamming point with the help of self-propulsion.
 The boundary is given by $\phi_J(\sigma^*, \sigma^*_a)$, which can be estimated from the critical value of the scaling variable $\tilde{x}^a_c = f(\sigma^*)C(\phi_J)g(\sigma_a^*)/(\phi_0 - \phi_J)$. With the help of linear fit of data shown in Fig.~\ref{fig3}(c), considering $C(\phi)$ a linear function $C(\phi) = 0.35 - 0.5 \phi$, we get
\begin{equation}
\phi_J(\sigma^*, \sigma_a^*) = \frac{\tilde{x}^a_c \phi_0 - 0.35 g(\sigma^*_a)f(\sigma^*)}{(\tilde{x}^a_c  - 0.5 g(\sigma^*_a)f(\sigma^*))} \text{.}
\label{Eq0}
\end{equation}    
Using Eq.~\ref{Eq0} and the aforementioned functional form of $f(\sigma^*)$ and $g(\sigma^*_a)$, for a fixed $\sigma_a^*$, one can get the $\sigma^*$ -- $\phi_J$ line as shown by the green solid line in Fig.~\ref{fig4}.
As $\sigma_a^*$ increases this line moves towards higher $\phi$, indicating that shear jamming occurs at larger volume fraction in the presence of self-propulsion. This is supported by the finding of Ref.~\cite{bhowmik2026PRE}.
However, since the $\eta_r$--$\sigma_a^*$ curve exhibits nonmonotonic dependence at large $\sigma_a^*$,
we are limited by a maximum practicable value of $\sigma_a^* \sim 150$.
Here the shear jamming region is considerably smaller than the activity-free case (Fig.~\ref{fig4}(b)), but unlike Ref.~\cite{MRamaswamyPRL2025} it does not vanish.
(Though earlier experiments do support a saturation of viscosity reduction under strong cross shear~\cite{LNCSC_PNAS_2016}.)

We can similarly estimate how the DST region alters with active stress.
This region is given by the condition $\frac{d \ln \eta_r}{d \ln \sigma^*} > 1$.
 Using the scaling relation $\eta_r \left(\phi_0 - \phi\right)^2 = \mathcal{F}(\tilde{x}^a)$ with $\tilde{x}^a = f(\sigma^*)C(\phi)g(\sigma_a^*)/(\phi_0 - \phi)$, we get
\begin{equation}
\frac{\sigma^{*}}{\mathcal{F}(\tilde{x}^a)}\frac{d \mathcal{F}(\tilde{x}^a)}{d \tilde{x}^a}\frac{df(\sigma^{*})}{d \sigma^{*}}\frac{C(\phi)g(\sigma_a^*)}{\phi_0 - \phi} > 1 \text{.}
\label{Eq1}
\end{equation}      
For large $\tilde{x}^a$, considering $\mathcal{F}(\tilde{x}^a) = (\tilde{x}^a_c - \tilde{x}^a)^{-\delta}$, we get
\begin{equation}
\frac{\delta \sigma^{*}}{(\tilde{x}^a_c - \tilde{x}^a)}\frac{df(\sigma^{*})}{d \sigma^{*}}\frac{C(\phi)g(\sigma_a^*)}{\phi_0 - \phi} > 1 \text{.}
\label{Eq2}
\end{equation}
Now, for a fixed $\sigma^*$, as $\sigma_a^*$ increases
$g(\sigma_a^*)$ decreases, which forces a decrease in $\phi_0 - \phi$ to satisfy the inequality in Eq.~\ref{Eq2}.
Thus with increasing active stress the boundary of the DST region moves towards $\phi_0$.
However, similar to the shear-jamming region, unlike OSP in Ref.~\cite{MRamaswamyPRL2025}, this shear thickening region will not vanish since we cannot push to the limit $g(\sigma_a^*) = 0$ by making $\sigma_a^* \rightarrow \infty$ since the viscosity starts increasing as we cross certain values of active stress due to activity induced diffusion. Quantitatively the phase boundary of the DST region (gray dashed line in Fig.~\ref{fig4}) is given by
\begin{equation}
\frac{\delta \sigma^{*}}{(\tilde{x}^a_c - \tilde{x}^a)}\frac{df(\sigma^{*})}{d \sigma^{*}}\frac{C(\phi)g(\sigma_a^*)}{\phi_0 - \phi} = 1 \text{,}
\label{Eq3}
\end{equation}
\begin{figure}
\includegraphics[width=0.4\textwidth]{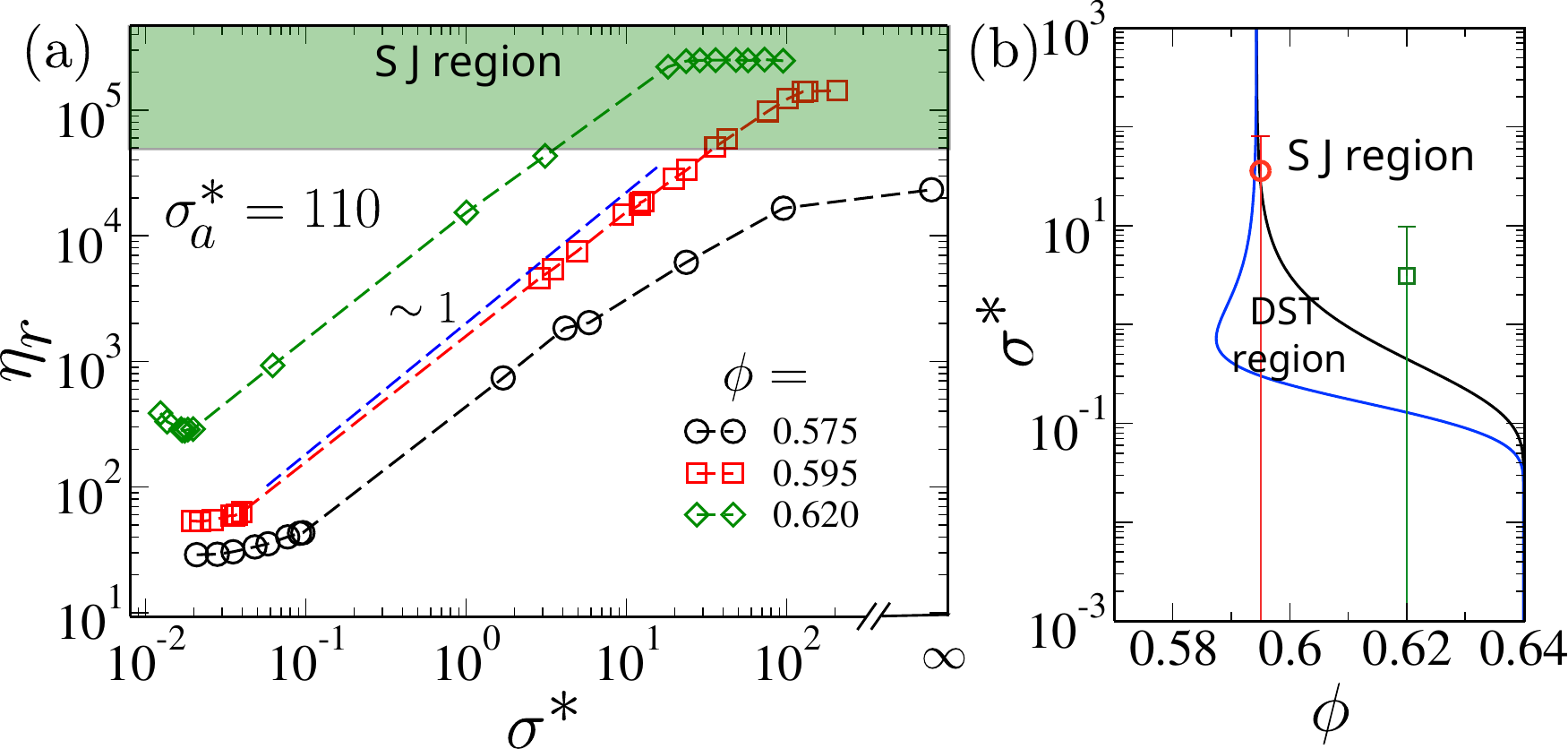}
\caption{(a) $\eta_r$ vs. $\sigma^*$ curves for three volume fractions, $\phi = 0.575$, 0.595, and 0.62, at an active stress $\sigma_a^* = 110$. The system exhibits DST (with $\frac{d\ln \eta_r}{d\ln \sigma^*} \geq 1$, indicated by the blue dashed line with slope 1) and SJ (data in the green region) for $\phi = 0.62$ and 0.595, but not for $\phi = 0.575$, as suggested by the phase diagram inferred from our scaling analysis shown in (b). The green circle and pink square with error bars in (b) indicate the shear-jamming points obtained from the simulations for $\phi = 0.595$ and $0.62$, respectively.}
\label{fig5}
\end{figure}
and can be estimated for a fixed $\sigma_a^*$ using the functional form of $f(\sigma^*)$ and $g(\sigma^*_a)$. Fig.~\ref{fig4}(b) shows the shear thickening region for a very large $\sigma^*_a = 150$.

We further extend our work to validate the scaling analysis through targeted tests of the accuracy of the inferred phase boundaries. We simulate our active system at a fixed $\sigma_a^* = 110$ for a set of volume fractions, $\phi = 0.575$, $0.595$, and $0.62$, above $\phi_J$. For a fixed $\phi$, these simulations exhibit continuous shear thickening, DST, and SJ, depending on the value of $\sigma^*$. We then compare these values of $\sigma^*$ with the inferred $\sigma^* - \phi$ phase diagram for $\sigma_a^* = 110$. The data are shown in Fig.~\ref{fig5}. The data for $\phi = 0.575$ show neither DST nor SJ with increasing $\sigma^*$, as suggested by the scaling analysis. However, $\phi = 0.595$ and $0.62$ exhibit both DST and SJ (shown in Fig.~\ref{fig5}(a)), in agreement with the estimated phase diagram. In our model with a repulsive force, we find it difficult to pinpoint the onset of DST. However, we compare the value of $\phi_J$ obtained from the simulations with the estimated value. Since the constituent particles in our simulations are not hard spheres, we define the SJ region using $\eta_r^J = 6 \times 10^4$, shown by the green region in Fig.~\ref{fig5}(a)~\cite{bhowmik2026PRE}. We then take the value of $\sigma^*$ that yields a viscosity closest to $\eta_r^J$ as the stress at which jamming occurs and compare it with the inferred value, as shown in Fig.~\ref{fig5}(b). Given the sensitivity of the data close to jamming, the estimated and observed values agree within the error bars. A direct numerical determination of the entire phase boundary over a range of $\sigma_a^*$ requires a vast amount of numerical simulation and is therefore not attempted here.
\paragraph{Conclusion.} 
We have studied tunable shear thickening in active dense suspensions of non-Brownian particles.
The degree of shear thickening can be tuned by varying a dimensionless active stress which characterises the ratio of the active stress to the stress due to external driving.
Self-propulsion of the particles reduces the number of frictional contacts,
similar to the mechanism observed in tunable shear thickening achieved by applying cyclic shear orthogonal to the primary flow direction.
A scaling framework that describes tunable shear thickening in the presence of OSP is used to explain the tunable shear thickening induced by self-propulsion,
leading to a successful collapse that supports this perspective as a unifying tool for suspension rheology.  In both OSP-like perturbations and isotropic activity, the additional particle motion suppresses frictional contacts, leading to similar rheological behaviour. Although Ref.~\cite{bhowmik2026PRE} shows how the viscosity of dense suspensions can be tuned by varying the active forcing, the study remains limited to the microscopic mechanism and the role of friction, whereas the scope of the present work goes beyond this. Here, we show how the various stress scales relevant to driving and self-propulsion compete and that the rheology is consistent with a previously established scaling framework. Using the resulting empirically calibrated scaling relation, we infer how activity alters the shear-thickening and jamming phase diagrams, and partially validate these inferences with targeted simulations at selected state points. It is important to note that the observed data collapse reflects only a rheological correspondence between OSP and active driving, rather than evidence of any universal similarity between the two mechanisms~\cite{RishabSmarajitNatPhy2025, Goswami2025}. For instance, in many systems above jamming the active driving might affect the rheology through coupling with long wavelength phonon modes which is not the case here~\cite{AntikNatComm2025}. Also, the observed tunable dethickening at low activity strength and thickening at high activity strength are specific to frictional systems. Tunable rheological responses have also been observed in many other soft active suspensions due to motility-induced phase separation~\cite{HurtmutSoftMatter} or the breaking of clusters due to activity~\cite{PNASThiining}. Therefore, activity does not always result in a similar rheological response; rather, the response is controlled by the microscopic details of the particles. 
Our study is conducted by varying the active force at a fixed persistence time;
therefore, the persistence time does not appear in the scaling variables.
However, we note that tunable shear thickening can also be achieved by varying the persistence time,
and indeed this may allow further optimisation of the $\sigma^*-\phi$ phase diagram, possibly eliminating the DST and SJ regions entirely.
In such cases, the scaling variables would need to be modified accordingly. Also, the data collapse does not include data at high active stress where viscosity starts increasing with $\sigma_a^*$. This thickening regime can be avoided by not driving the system at high $\sigma_a^*$ or by choosing a different mode of self-propulsion such as active torque which contributes less to the translational motion. In this study, all particles are considered active. We note that the observed phenomena and the scaling framework should remain qualitatively unchanged when only a fraction of the particles are active. A reduction in the fraction of active particles can be compensated for by an increase in the magnitude of the self-propulsion force~\cite{bhowmik2026PRE}. Moreover, this work considers only active particles modelled as run-and-tumble particles with sliding friction. Other forms of activity, such as active Brownian particles or active torques and inclusion of rolling friction are also expected to produce similar dethickening behaviour. A detailed investigation of the effects of persistence time and different models of activity is left for future work.
\paragraph{Acknowledgement.}
B.P.B. acknowledges support from the Department of Science and Technology (DST), Government of India, under the INSPIRE Faculty Fellowship scheme (Grant No. DST/INSPIRE/IFF CALL-24/IFA24-PH 322), and the PARAM Brahma computing facility at IISER Pune under the National Supercomputing Mission, Government of India.
We thank Anna Barth and Itai Cohen for discussions.

\bibliography{ALL1}
\end{document}